\documentclass[12pt,a4paper]{article}
\pdfoutput=1

\usepackage{amsfonts,amsthm}
\usepackage{jcappub}

\newcommand{\beq}{\begin{equation}}
\newcommand{\eeq}{\end{equation}}

\title{Starobinsky-Type Inflation from $\alpha'$ - Corrections}

\author[1]{Benedict J. Broy}
\author[1]{David Ciupke}
\author[2]{Francisco G. Pedro}
\author[1]{Alexander Westphal}

\affiliation[1]{\emph{Deutsches Elektronen-Synchrotron DESY, Theory Group, 22603 Hamburg, Germany}}
\affiliation[2]{\emph{Departamento de F\'{\i}sica Te\'orica and Instituto de F\'{\i}sica Te\'orica  UAM-CSIC, Universidad Aut\'onoma de Madrid, Cantoblanco, 28049 Madrid, Spain}}

\emailAdd{benedict.broy@desy.de}
\emailAdd{david.ciupke@desy.de}
\emailAdd{francisco.pedro@csic.es}
\emailAdd{alexander.westphal@desy.de}

\abstract{Working in the Large Volume Scenario (LVS) of IIB Calabi-Yau flux compactifications, we construct inflationary models from recently computed higher derivative $(\alpha')^3$-corrections. Inflation is driven by a K\"ahler modulus whose potential arises from the aforementioned corrections, while we use the inclusion of string loop effects just to ensure the existence of a graceful exit when necessary. The effective inflaton potential takes a Starobinsky-type form $V=V_0(1-e^{-\nu\phi})^2$, where we obtain one set-up with $\nu=-1/\sqrt{3}$ and one with $\nu=2/\sqrt{3}$ corresponding to inflation occurring for increasing or decreasing $\phi$ respectively. The inflationary observables are thus in perfect agreement with PLANCK, while the two scenarios remain observationally distinguishable via slightly varying predictions for the tensor-to-scalar ratio $r$. Both set-ups yield $r\simeq (2\ldots 7)\,\times 10^{-3}$. They hence realise inflation with moderately large fields $\left(\Delta\phi\sim 6\thinspace M_{Pl}\right)$ without saturating the Lyth bound. Control over higher corrections relies in part on tuning underlying microscopic parameters, and in part on intrinsic suppressions. The intrinsic part of control arises as a leftover from an approximate effective shift symmetry at parametrically large volume.}

\arxivnumber{1509.00024}

\begin{document}

\maketitle
\flushbottom




\section{Introduction} 

Inflationary cosmology constitutes a rather successful paradigm for solving the conventional hot big bang initial condition problems. The simplest models of inflation create the very early exponentially accelerated expansion via vacuum energy domination of a single slowly rolling scalar field. While inflation enjoys by now robust support from probes of the cosmic microwave background (CMB) such as WMAP, Planck, and BICEP2/Keck Array \cite{Bennett:2012zja,Ade:2015lrj,Ade:2015tva}, it is in need of UV completion due to its intrinsic sensitivity to higher order corrections to the relevant couplings in the scalar Lagrangian describing the slow-roll process.

One of the historically earliest models of slow-roll inflation arises in the Starobinsky model~\cite{Starobinsky:1980te} (for the simplified $R+R^2$ form see also~\cite{Barrow:1988xh}) by changing the Einstein-Hilbert gravitational action to a form 
\beq\label{SRsquared}
S=\int{\rm d}^4x\sqrt{-g}\,\frac{M_{Pl}^2}{2}\,\left(R+\frac{1}{6M^2}R^2\right)\quad.
\eeq
This is conformally equivalent to a canonically normalised scalar field  with an exponentially flat `plateau' potential
\beq
V=V_0\,\left(1-e^{-\sqrt{2/3}\phi}\right)^2
\eeq
(and further to a quartic potential for a canonically normalised, but non-minimally coupled scalar field in `Jordan frame', as e.g. in Higgs inflation~\cite{Bezrukov:2007ep}). Models of this `plateau' type have been studied extensively in recent years where they have been found to arise as attractor limits of a large variety of different inflaton potentials in the context of Jordan frame supergravity~\cite{Kallosh:2013xya,Kallosh:2013hoa,Buchmuller:2013zfa,Ferrara:2013rsa,Kallosh:2013maa,Kallosh:2013tua,Kallosh:2013yoa,Kallosh:2014rga,Galante:2014ifa,Kallosh:2015zsa,Carrasco:2015uma,Carrasco:2015pla,Broy:2015qna}. Moreover, it predicts inflationary observables $n_s=1-2/N\simeq 0.97$ and $r=12/N^2\simeq 0.003$ well in agreement with the recent CMB data.

Note, that COBE normalisation of the curvature perturbation requires $V_0\sim M^2M_{Pl}^2\sim 10^{-10}M_{Pl}^4$ ). This tells us that $M\sim 10^{-5}M_{Pl}$ is small, and thus the $R^2$ appears strongly enhanced. Hence, positing literally just the semi-infinite plateau potential or, equivalently, the above modified gravitational action would correspond to defining the UV theory of gravity by eq.~\eqref{SRsquared}.

However, once we look at an actual candidate for a UV theory of quantum gravity, string theory (or even just the supergravity embeddings of Starobinsky-type potentials mentioned above) tells us that the simple UV structure of gravity assumed in eq.~\eqref{SRsquared} does not hold. Already the first extant string or supergravity embeddings of such plateau-type moduli inflation models (such as fibre inflation~\cite{Cicoli:2008gp} or poly-instanton inflation~\cite{Cicoli:2011ct}) show that the pure half-infinitely extended plateau potential acquires a whole series of either exponentially rising or falling corrections at a finite distance in field space which limit the width of the inflationary plateau. Clearly, even when restricting the description of these potentials to the pure Einstein gravity + inflaton scalar field sector, a 4D conformal transformation of these models into the Starobinsky-like frame of eq.~\eqref{SRsquared} would show the appearance of $R^n,n\geq 3$ terms (besides all the other non-$R^n$-type higher curvature terms descending directly from string compactification), see e.g.~\cite{Broy:2014xwa}, which invalidate the assumed UV form of eq.~\eqref{SRsquared}.

Existence of an extended inflationary plateau of width $\Delta\phi\sim 5\,M_{Pl}$ necessary for about 60 e-folds of slow-roll inflation requires an all-order suppression of the prefactors of higher-order rising or falling exponential corrections. Hence, this motivates constructing such exponential plateau-type potentials with plateaus of finite width limited by exponentially rising of falling corrections in UV completions like string theory.

Compactifications of type IIB string theory on Calabi-Yau orientifolds with 3-form flux fixing the complex structure moduli and the axio-dilaton, and a combination of non-perturbative superpotential corrections and ${\cal O}(\alpha'^3)$ corrections to the K\"ahler potential allow for a landscape of moduli-free metastable dS vacua with certain classes of inflationary regions arising from flat potential regions for the K\"ahler moduli. In a subclass of these models called the Large Volume Scenario (LVS) compactification on a K3-fibred Calabi-Yau produces a K\"ahler moduli scalar potential which is flat at leading order in the LVS mechanism for the volume modulus $\tau_1$ of the K3-fibre. In `fibre inflation'~\cite{Cicoli:2008gp} the generic form of string loop corrections to the K\"ahler potential~\cite{Cicoli:2007xp,Cicoli:2008va} produces a scalar potential with a minimum for the fibre modulus, and upon tuning the string loop coefficients via their complex structure moduli dependence and under certain assumptions about the higher string-loop contributions the 1-loop terms generate an extended exponential plateau potential driving inflation as well. The exponential form the potential terms arises in these models from canonically normalising the moduli kinetic terms. A different but closely related model is that of poly-instanton inflation~\cite{Cicoli:2011ct} where instantonic corrections to instantons are responsible for the genesis of the potential for the fibre modulus. In this case the potential features double exponentials, which make it flatter than Starobinsky-like models, and imply a lower tensor signal. In both of these fibre inflation setups the width of the inflationary plateau is always limited by the appearance of rising exponential terms in the scalar potential from some of the string loop corrections.

In this paper we report on an improved construction of Starobinsky-type plateau potentials in type IIB string compactifications based on the original setup of fibre inflation in~\cite{Cicoli:2008gp}. This construction uses the recent discovery~\cite{Ciupke:2015msa} of a class of four-superspace-derivative operator contributions to the scalar moduli potential arising from the 10D ${\cal O}(\alpha'^3) R^4$ corrections. We can arrange for a trans-Planckian inflationary plateau for the fibre modulus to arise \emph{only} from the scalar potential induced by this new ${\cal O}(\alpha'^3)$-contribution and the well known ${\cal O}(\alpha'^3)$-correction the volume moduli K\"ahler potential, with the string loop corrections relegated to the tasks of just providing post-inflationary minima for the fibre modulus. 

This provides us with a modicum of control over higher-order corrections, since the inflationary plateau arises from $\alpha'$-corrections only, and the $g_s$-suppression of the string loops guarantees an all-order protection from the loops once the 1-loop contributions are small. Control over even higher-order $\alpha'$-corrections in turn arises from the suppression of these terms at large stabilised volume. This leaves the remaining amount of necessary tuning to appear in the (complex structure moduli VEV dependent) string loop coefficients, which need to be subdominant to the $\alpha'$-corrections on the inflationary plateau while generating a minimum for the inflaton fibre modulus away from the plateau. Effectively, the smallish-$g_s$ based decoupling of the string loops from the inflationary plateau, and the large-volume control of successively higher-order $\alpha'$-corrections constitutes a leftover from the approximate K\"ahler moduli shift symmetry at parametrically large volume.

Dividing the tasks of generating a minimum for the fibre modulus and generating its inflationary plateau between the string loop and $\alpha'$-corrections this way instead of having one type corrections doing everything leads in addition to a mini-landscape of inflationary regimes and minima -- namely it is possible to get inflation rolling both \emph{to the right} and \emph{to the left}. The inflationary potential then takes a form following the discussion above which can be summarised as
\beq
V=\left\{\begin{array}{ll} V_0\,\left(1-e^{-\nu\phi}\right)^2+\varepsilon^2\,e^{\frac{\nu}{2}\phi}+\ldots & \textrm{rolling to the left}\\ & \\ V_0\,\left(1-e^{\nu\phi}\right)^2+\varepsilon^2\,e^{-2\nu\phi}+\ldots & \textrm{rolling to the right} \end{array} \right. \quad.
\eeq
A successful construction arranges for $\varepsilon\ll 1$ ensuring the appearance of a finite-width inflationary plateau. For the simple toy K3-fibred geometries we are looking at, we find that the structure of the four-derivative $\alpha'$-correction~\cite{Ciupke:2015msa} generating the plateau region of the potential leads to values for $\nu=2/\sqrt 3$, $\nu=1/\sqrt 3$ differing between inflation rolling to the left, and to the right, respectively. The predictions for the inflationary observables hence split as well into $n_s=1-2/N\simeq 0.97$, and $r\simeq 0.002$ vs $r\simeq 0.007$, respectively.

The rest of the paper is structured as follows. We first review the framework of LVS and recall the recent $(\alpha')^3$-corrections to the 4D effective action of type IIB Calabi-Yau-orientifold compactifications with background fluxes in section~\ref{LVSIIBandCorrs}. In section~\ref{Construct} we describe the toy 3-parameter setup of a K3-fibred Calabi-Yau and its K\"ahler moduli scalar potential arising from the combination of a standard LVS-type stabilisation of two of the three volume modulus and the fibre modulus acquiring a plateau inflation scalar potential with minimum from the combined effect of the string loop and two- and four-derivative ${\cal O}(\alpha'^3)$ contributions to the scalar potential. We discuss the various left- and right-rolling branches of this `breathing fibre inflation', and then move to the extraction of the inflationary observables in section~\ref{observables}. We finally conclude in section~\ref{discussion}.
 
\section{Large Volume Limit of IIB Flux compactifications}\label{LVSIIBandCorrs}

\subsection{The Large Volume Scenario}

We will begin by briefly reviewing the framework of LVS. Our starting point is the large volume limit of the low-energy 4D $\mathcal{N}=1$ - effective action of type IIB Calabi-Yau-orientifold compactifications with background fluxes \cite{Becker:2002nn, Grimm:2004uq} including the leading order $(\alpha')^3$-corrections to the bulk fields \cite{Becker:2002nn} and string-loop corrections \cite{Berg:2005ja, Berg:2007wt}. Furthermore, we consider non-perturbative corrections descending from gaugino condensation on wrapped D7-branes or from Euclidean wrapped D3-branes. The background fluxes admit supersymmetric minima for the dilaton as well as complex structure moduli at tree level. After replacing these fields with their respective minima in the effective Lagrangian, the theory is subject to the following K\"ahler and superpotential
\begin{equation}\begin{aligned}\label{K_W}
 K(T + \bar{T}) &= \ln (g_s) - 2 \ln (\mathcal{V}+\tfrac{1}{2}\hat\xi) + \delta K^{KK}_{g_s} + \delta K^{W}_{g_s} \\
 W(T) &= W_0  + \sum_i A_i \mathrm{e}^{-a_i T_i} \equiv \frac{1}{\sqrt{2}} \mathrm{e}^{\langle K_{cs} \rangle / 2} \left( \tilde{W}_0  + \sum_i \tilde{A}_i \mathrm{e}^{-a_i T_i} \right) \ .
\end{aligned}\end{equation}
Here $K_{cs}$ denotes the K\"ahler potential of the complex structure moduli, $g_s$ the string coupling and $\tilde{W}_0$ is the Gukov-Vafa-Witten superpotential evaluated at the supersymmetric minimum.\footnote{Here we have absorbed the term $\langle K_{cs} \rangle-\ln(2)$ into the superpotential by means of a K\"ahler transformation.} The non-perturbative corrections from wrapped D-branes induce a term in the superpotential. Here $\tilde{A}_i$ and $a_i$ can be regarded as constants, that are related to the specifics of the mechanism which generates these non-perturbative contributions. In order to be able to neglect higher instanton corrections, it is necessary to ensure that $a_i T_i \gg 1$. Moreover, the total volume modulus reads
\begin{equation}
 \mathcal{V} = \tfrac{1}{6} k_{ijk} t^i t^j t^k \ ,
\end{equation}
where $k_{ijk}$ denote triple intersection numbers and $t^i$ two-cycle volumes. The latter are implicitly related to the real parts of the K\"ahler coordinates, which are given by the dual four-cycle volumes via
\begin{equation}
 \frac{1}{2}(T_i + \bar{T}_i) \equiv \tau_i = \frac{\partial \mathcal{V}}{\partial t^i} = \tfrac{1}{2} k_{ijk} t^j t^k \ .
\end{equation}
Furthermore, $\hat\xi$ parametrises the leading order $(\alpha')^3$-corrections and reads \cite{Becker:2002nn}
\begin{equation}
 \hat\xi = - \frac{(\alpha')^3 \zeta(3) \chi(M_3)}{2(2\pi)^3 g_s^{3/2}} \ ,
\end{equation}
where $\chi(M_3)$ denotes the Euler number of the threefold. Lastly, the K\"ahler potential includes string-loop corrections which are induced from the exchange of closed strings carrying Kaluza-Klein momentum as well as from the exchange of winding strings. Their general form was inferred in \cite{Berg:2007wt} to be
\begin{equation}\begin{aligned}
\label{eq:string_loop_corrections}
 \delta K^{KK}_{(g_s)} & \sim g_s \sum_{i=1}^{h^{1,1}} \frac{C_i^{KK} (a_{ij}t^j)}{\mathcal{V}} \ , \qquad
 \delta K^{W}_{(g_s)} \sim \sum_{i=1}^{h^{1,1}} \frac{C_i^W (a_{ij}t^j)^{-1}}{\mathcal{V}} \ .
\end{aligned}\end{equation}
Here $C_i^{KK}$ and $C_i^W$ can be regarded as constants, which depend on the complex structure moduli at the minimum, and $a_{ij}$ are combinatorial constants. For the particular example of $T^6/(\mathbb{Z}_2 \times \mathbb{Z}_2)$ the explicit form of $C_i^{KK}$ and $C_i^W$ was computed in \cite{Berg:2005ja}. For this example it was shown in \cite{Berg:2005yu} that as long as the complex structure moduli are not stabilised large, and note that stabilising the complex structure moduli at large values would include a certain amount of fine-tuning, one finds that roughly
\begin{equation}\label{C_gs_estimate}
 C_i^{KK} \simeq C_i^W \simeq \frac{1}{128 \pi^4} \ ,
\end{equation}
which in the following will serve as a ballpark estimate of these constants. 
The scalar potential derived from \eqref{K_W} can be expanded in the large volume limit in inverse powers of $\mathcal{V}$. The leading order contributions from the individual terms \eqref{K_W} read
\begin{equation}
 V = V^{LVS} + \delta V_{(g_s)} \equiv V_{np} + V_{\alpha'} + \delta V_{(g_s)} \ ,
\end{equation}
where \cite{Cicoli:2007xp, Cicoli:2008va}
\begin{equation}\begin{aligned}\label{gen_potential}
 V_{np} &= \mathrm{e}^K K_0^{i \bar{j}} \left( a_i a_j A_i \bar{A}_{\bar{j}} \mathrm{e}^{-(a_i T_i + a_j \bar{T}_{\bar{j}})}  - a_i A_i \mathrm{e}^{-a_i T_i} \bar{W} K_{0,\bar{j}} - a_j \bar{A}_{\bar{j}} \mathrm{e}^{-a_j \bar{T}_{\bar{j}}} W K_{0,i} \right) \\
 V_{\alpha'} &= g_s \frac{3\hat\xi \lvert W_0 \lvert^2}{4 \mathcal{V}^3} \ , \qquad \delta V_{(g_s)} = \sum_i \frac{g_s \lvert W_0 \lvert^2}{\mathcal{V}^2} \left[ g_s^2 (C^{KK}_i)^2 K_{(0),ii} - 2 \delta K^{W}_{(g_s), t^i} \right] \ .
\end{aligned}\end{equation}
Here $K_{0} = -2 \ln \mathcal{V}$ and we have assumed that the D-branes only wrap basis four-cycles. To achieve a large stabilised volume additional assumptions regarding the triple intersection numbers have to be made. In the original model in \cite{Balasubramanian:2005zx} the volume was assumed to be controlled by a single four-cycle and that furthermore a blow-up cycle exists, for which non-perturbative corrections to $W$ are present. Then $V^{LVS}$ has a controlled minimum with exponentially large volume. The string-loop correction $\delta V_{(g_s)}$ is particularly interesting for those compactification-geometries for which $V^{LVS}$ has a flat direction or no minimum at large volume exists. A particular case, in which $V^{LVS}$ has an exact flat direction was studied in \cite{Cicoli:2008va, Cicoli:2008gp}.
\subsection{Four-Superspace-Derivative Correction}
Besides the $(\alpha')^3$-correction to the K\"ahler potential in \eqref{K_W} further $(\alpha')^3$-corrections are expected to enter the 4D effective action of type IIB Calabi-Yau-orientifold compactifications with background fluxes. It was argued recently in \cite{Ciupke:2015msa} that these corrections have to be described off-shell via higher-derivative terms in $\mathcal{N}=1$ - superspace. Moreover, in \cite{Ciupke:2015msa} the leading order operator in the higher-superspace-derivative expansion was computed by KK-reducing $(\alpha')^3$-corrections in 10D. At order $(\alpha')^3$ this operator contributes four-derivative and additional two-derivative-terms as well as a new term in the scalar potential to the effective action of the K\"ahler moduli. Even though the derivative-terms are not known exactly, the functional form of the correction to the scalar potential could be inferred by using the no-scale property of $K_{0}$ and reads\footnote{Strictly speaking this formula is only valid for models without non-perturbative corrections to $W$. However, we will consider models where $V_{(1)}$ is the relevant potential for those fields which are flat directions of $V^{LVS}$ and for which non-perturbative corrections are either absent or negligible so that indeed the form of $V_{(1)}$ is correct. Blow-ups and the volume modulus are stabilised by $V^{LVS}$ and are approximately unaffected by $V_{(1)}$.}
\begin{equation}\label{correction}
V_{(1)}= - g_s^2 \hat\lambda\frac{{|W_0|}^4}{\mathcal{V}^4} \Pi_i t^i \thinspace,
\end{equation}
where $\hat\lambda = \lambda (\alpha')^3 g_s^{-3/2}$ with $\lambda$ being an undetermined combinatorial constant. From the 10D origin of the correction one can infer the rough estimate
\begin{equation}\label{estimate_lambda}
\lvert \hat\lambda \lvert \, \simeq \Bigl\lvert \frac{\hat\xi}{\chi(M_3)} \Bigr\lvert \ , \qquad \text{and hence} \qquad  \lvert \lambda \lvert \, \simeq \frac{\zeta(3)}{2(2\pi)^3} \ .
\end{equation}
 Furthermore, the $\Pi_i$ in \eqref{correction} are integer numbers encoding the topological information of the second Chern class $c_2(M_3)$. More precisely, upon choosing $\hat{D}_i$ as a basis of harmonic $(1,1)$-forms on $M_3$, such that the K\"ahler form reads
\begin{equation}\label{J}
 J = \sum_{i=1}^{h^{1,1}} \hat{D}_i t^i \ ,
\end{equation}
we have
\begin{equation}\label{def_pi}
 \Pi_i = \int_{M_3} c_2 \wedge \hat{D}_i \ .
\end{equation}
If the two-cycles were chosen such that \eqref{J} holds, then they span the K\"ahler-cone and automatically $t^i \geq 0$ independently for each $i=1,\dots,h^{1,1}$. This in turn implies that $\Pi_i \geq 0$ for each $i=1,\dots,h^{1,1}$. With respect to an arbitrary choice of two-cycles the numbers $\Pi_i$ can have both signs.
\section{Towards an Inflationary Regime}\label{Construct}
\subsection{Volume Stabilisation for $K3$-Fibered Threefold}
We will consider the compactification geometry investigated in \cite{Cicoli:2008va, Cicoli:2008gp} for which a large-volume scenario with a flat direction exists (that is without including $V_{(1)}$ or $\delta V_{(g_s)}$). More precisely, this geometry can be regarded as ``adding a blow-up cycle'' to the $K3$-fibered threefold $\mathbb{CP}^4 [1,1,2,2,6]$ which has $h^{1,1}=2$. The volume in terms of the four-cycles is assumed to be of the form
\begin{equation}\label{Vol_fibre}
\mathcal V=\alpha\left(\sqrt{\tau_1}\tau_2 - \gamma \tau_3^{3/2} \right) \ ,
\end{equation}
where $\tau_1$ is associated with the volume of the $K3$-fibre, $\tau_2$ controls the overall volume and $\tau_3$ denotes the blow-up. It is possible to find intersection numbers, which yield a volume as in \eqref{Vol_fibre} \cite{Cicoli:2008gp}. For instance choosing
\begin{equation}
 \mathcal V = \lambda_1 t^1 (t^2)^2+\lambda_3 (t^3)^3 \ ,
\end{equation}
we find
\begin{equation}\label{4cycle_2cycle}
\tau_1 = \lambda_1 (t^2)^2\thinspace, \quad \tau_2 = 2\lambda_1 t^1 t^2 \thinspace, \quad \tau_3 = 3 \lambda_3 (t^3)^2 \ .
\end{equation}
We will now assume that we are not in the K\"ahler cone, but that we have chosen a basis such that $t^3 \leq 0$. Inverting \eqref{4cycle_2cycle} yields
\begin{equation}\label{2cycle_4cycle}
t^1=\frac{\tau_2}{2\sqrt{\lambda_1 \tau_1}} \thinspace, \quad t^2= \sqrt{\frac{\tau_1}{\lambda_1}}  \thinspace, \quad t^3 = -\sqrt{\frac{\tau_3}{3\lambda_3}} \ .
\end{equation}
Hence, one indeed obtains \eqref{Vol_fibre} with $\alpha = \tfrac{1}{2}\lambda_1^{-1/2}$ and $\gamma = \sqrt{\tfrac{4}{27}} \sqrt{\lambda_1} \lambda_3^{-3/2}$. Let us now briefly summarise the volume stabilisation for \eqref{Vol_fibre} following \cite{Cicoli:2008gp}. Firstly, the limit of large volume requires $\tau_1 , \tau_2 \gg \tau_3$ and, thus, the only relevant terms in the superpotential are given by
\begin{equation}
 W(T) \simeq W_0 +  A_3 \mathrm{e}^{-a_3 T_3} \ .
\end{equation}
Calculating the K\"ahler metric and minimising the axionic component of $T_3$ one obtains \cite{Cicoli:2008gp} 
\begin{equation}
 V^{LVS}(\mathcal{V},\tau_3) = g_s \left[\frac{8a_3^2 A_3^2}{3\alpha \gamma} \frac{\sqrt{\tau_3}}{\mathcal{V}} \mathrm{e}^{-2 a_3 \tau_3} - 4 W_0 a_3 A_3 \frac{\tau_3}{\mathcal{V}^2} \mathrm{e}^{- a_3 \tau_3} + \frac{3\hat\xi W_0^2}{4\mathcal{V}^3} \right]\ .
\end{equation}
The above potential indeed does not depend on $\tau_1$, which is, hence, left as a flat direction. $V^{LVS}$ has a minimum at exponentially large volume given by
\begin{equation}\label{LVS_min}
 \langle \tau_3 \rangle = \left( \frac{\hat\xi}{2 \alpha \gamma} \right)^{2/3}\ , \quad \langle \mathcal{V} \rangle = \frac{3 \alpha \gamma}{4 a_3 A_3} W_0 \sqrt{\langle \tau_3 \rangle} \mathrm{e}^{a_3 \langle \tau_3 \rangle} \ .
\end{equation}
This minimum is expected to be perturbatively stable as long as no unnaturally large coefficients of higher-order operators occur \cite{Conlon:2005ki}. In the following we assume that $\tau_3$ and $\mathcal{V}$ are sitting at the minimum \eqref{LVS_min} and turn our attention fully to the flat direction $\tau_1$, for which additional perturbative corrections, that is both $\delta V_{(g_s)}$ as well as $V_{(1)}$, are important. Moreover, as the size of these corrections is sub-leading compared to $V^{LVS}$ we naively expect $\tau_1$ to be the lightest modulus and, hence, $\tau_1$ possibly constitutes a candidate to drive inflation.
\subsection{Perturbative Corrections and Inflation}
We begin the analysis of additional corrections to the scalar potential by the higher-derivative $(\alpha')^3$-correction given in \eqref{correction}.
For the geometry in defined in \eqref{Vol_fibre} and \eqref{2cycle_4cycle} $V_{(1)}$ reads
\begin{equation}\label{V1correction}
V_{(1)} \simeq -g_s^2 \hat\lambda\frac{{|W_0|}^4}{\mathcal V^4}\left(\Pi_1 \frac{\mathcal V}{\tau_1}+\Pi_2 \lambda_1^{-1/2}\sqrt{\tau_1}\right) \ .
\end{equation}
Note that we neglected all $\tau_3$-dependent terms in \ref{V1correction}. In what follows we will ignore such terms as they play no role in the phenomenology we are interested in here.

After replacing $\mathcal V$ and $\tau_3$ with their values at the LVS minimum \eqref{LVS_min} the scalar potential, excluding the string-loop corrections for the moment, is of the form
\begin{align}
V(\tau_1)= V^{LVS}(\langle \tau_3 \rangle, \langle \mathcal{V} \rangle) - g_s^2 \hat\lambda\frac{{|W_0|}^4}{\langle \mathcal{V} \rangle^4}\left( \Pi_1 \frac{\langle \mathcal{V} \rangle}{\tau_1}+\Pi_2 \lambda_1^{-1/2}\sqrt{\tau_1}\right) \ .
\label{eq:Vtot}
\end{align}
In the case $\Pi_1, \Pi_2>0$ and $\hat\lambda<0$ this potential features a minimum. Slow roll inflation can in principle occur around this minimum. It is convenient to discuss inflation in terms of a canonically normalised field. The normalisation can be performed following \cite{Cicoli:2008gp}. After replacing $\tau_2 = \tau_2(\tau_1,\mathcal{V},\tau_3)$ via \eqref{Vol_fibre} in the Lagrangian, the kinetic terms read
\begin{equation}
\mathcal{L} \supset -\frac{3}{8\tau_1^2} (\partial_\mu \tau_1 \partial^\mu \tau_1) + \frac{1}{2\tau_1 \mathcal{V}}  (\partial_\mu \tau_1 \partial^\mu \mathcal{V}) - \frac{1}{2\mathcal{V}^2} (\partial_\mu \mathcal{V} \partial^\mu \mathcal{V}) + \dots
\end{equation}
Thus, the canonically normalised inflaton $\varphi$ is related to $\tau_1$ as follows
\begin{equation}\label{canonical_normalization}
 \tau_1=e^{2 \varphi /\sqrt{3}} \ .
\end{equation}
In this formula $\tau_3$-dependent corrections were neglected. In \cite{Cicoli:2008gp} these have already been argued to be negligible. Moreover, the kinetic terms obtain $(\alpha')^3$-corrections both from the $\hat\xi$-term in the K\"ahler potential in \eqref{K_W} as well as the higher superspace-derivative correction \cite{Ciupke:2015msa}. For the former correction the canonical normalisation is altered schematically at leading order in $\alpha'$ as
\begin{equation}
 \tau_1 \simeq e^{2 \varphi /\sqrt{3}} \left( 1 + \hat\xi \frac{\varphi}{\mathcal{V}} + \mathcal{O}(\alpha'^6) \right) \ .
\end{equation}
Such corrections are negligible as long as $\varphi \ll \mathcal{V} / \hat\xi$. Unfortunately the higher superspace-derivative correction to the two-derivative term is not exactly known. Nevertheless, we can give an estimate based on the approximate correction \cite{Ciupke:2015msa}
\begin{equation}
 \mathcal{L} \supset g_s \hat\lambda \lvert W_0 \lvert^2 \frac{\Pi_l t^l }{\mathcal{V}^2} K_{0,i} K_{0,j}\, (\partial_\mu \tau_i \,\partial^\mu \tau_j) \ .
\end{equation}
Using \eqref{Vol_fibre} and \eqref{2cycle_4cycle} we again calculate the leading order correction to the canonical normalisation, this yields
\begin{equation}
 \tau_1 \simeq e^{2 \varphi /\sqrt{3}} + g_s \hat\lambda \lvert W_0 \lvert^2 \left(\frac{\Pi_1}{\mathcal{V}} + \frac{\Pi_2}{\mathcal{V}^2} e^{\sqrt{3} \varphi } \right) + \mathcal{O}(\alpha'^6) \ .
\end{equation}
The corrections are again numerically suppressed due to the appearance of $\hat\lambda$ and inverse volume factors. The last correction will, however, become important for $\varphi \gg 1$. In the following we will be interested in an intermediate $\varphi$-regime, where we can safely neglect the above corrections.

Using \eqref{canonical_normalization} we can rewrite the scalar potential as
\begin{equation}\label{firstAttempt}
V(\varphi)=V^{LVS}(\langle \tau_3 \rangle, \langle \mathcal{V} \rangle)-g_s^2 \hat\lambda\frac{{|W_0|}^4}{\langle \mathcal{V} \rangle^4}\left( \Pi_1 \langle \mathcal{V} \rangle e^{-2/\sqrt{3}\thinspace\varphi}+\Pi_2 \lambda_1^{-1/2}e^{\varphi/\sqrt{3}}\right).
\end{equation}
Inflation driven by a single rising exponential $V\sim e^{\nu \varphi}$ has $n_s=1-\nu^2$  and $r=8 \nu^2$ and is therefore well outside the preferred region in the $(n_s,r)$ plane \cite{Ade:2015lrj} with $\nu=1/\sqrt{3}$ or $\nu=-2/\sqrt{3}$. So in this case even though the lightest modulus, $\tau_1$,  is stabilised by the $(\alpha')^3$-corrections and the resulting potential can lead to a long lasting period of inflation, its observational signatures are essentially ruled out by current data.

However, while $\tau_1$ is the lightest modulus which renders the leading-order dynamics single-field, the remaining heavier moduli will eventually backreact due to the inflationary vacuum energy built up by displacing $\tau_1$. This eventual backreaction will try to lower the overall vacuum energy during inflation and will hence lead to flattening~\cite{Dong:2010in} of the effective inflaton potential. We leave it for future work to determine if this effect applied to exponential scalar potentials is strong enough to lower $r$ and shift $n_s$ sufficiently to acquire observational compatibility for exponential field dependence with ${\cal O}(1)$ coefficients in the exponents.

\begin{figure}[h!]
	\centering
	\begin{minipage}[b]{0.48\linewidth}
	\centering
	\includegraphics[width=1\textwidth]{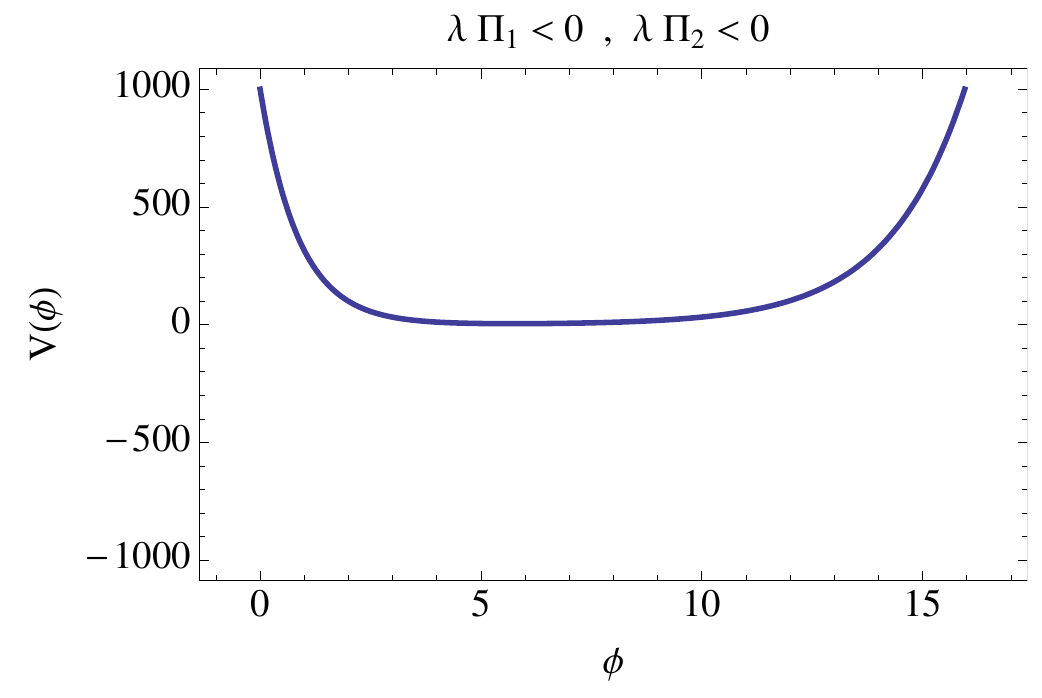}
    \end{minipage}
	\hspace{0.05cm}
	\begin{minipage}[b]{0.48\linewidth}
	\centering
	\includegraphics[width=1\textwidth]{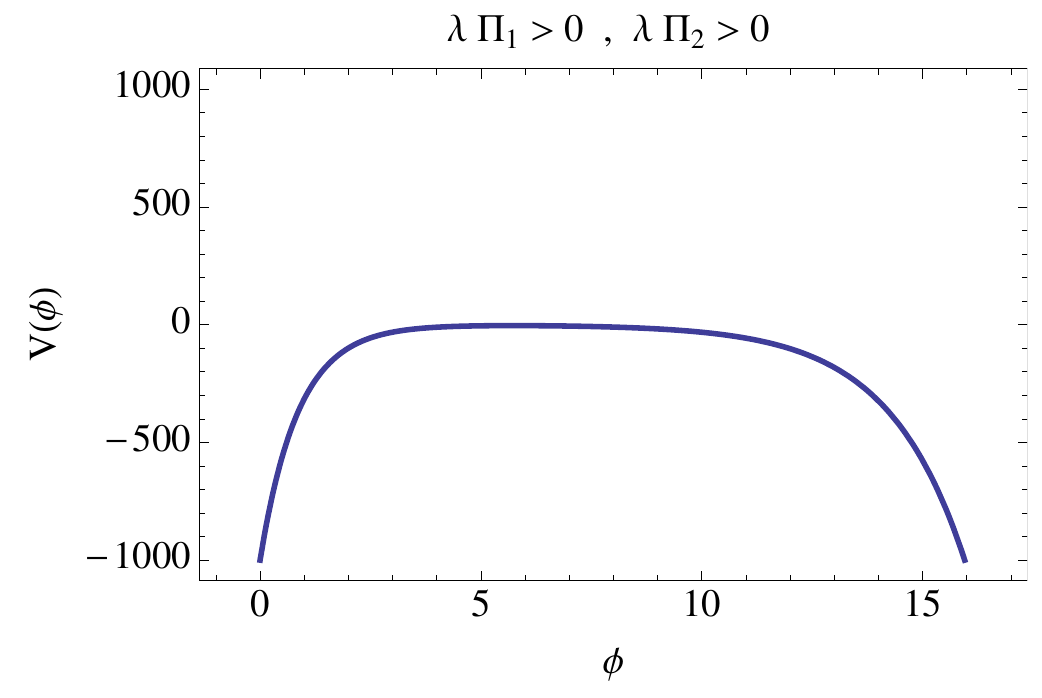}
	\end{minipage}
	\hspace{0.05cm}
	\centering
	\vspace{0.05cm}
	\begin{minipage}[b]{0.48\linewidth}
	\centering
	\includegraphics[width=1\textwidth]{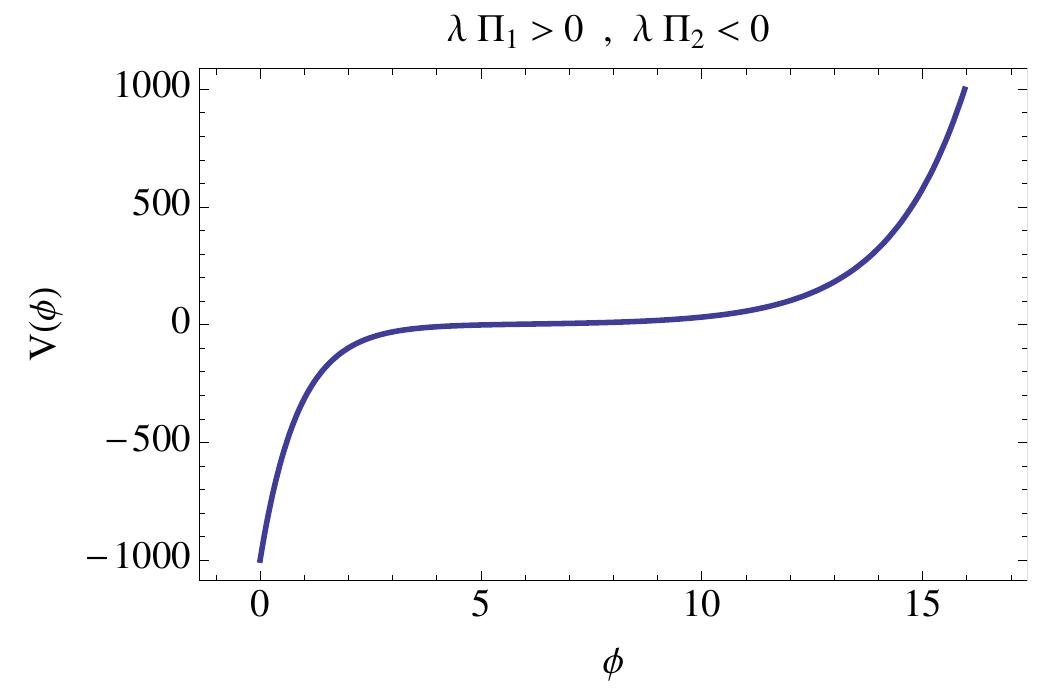}
    \end{minipage}
	\hspace{0.05cm}
	\begin{minipage}[b]{0.48\linewidth}
	\centering
	\includegraphics[width=1\textwidth]{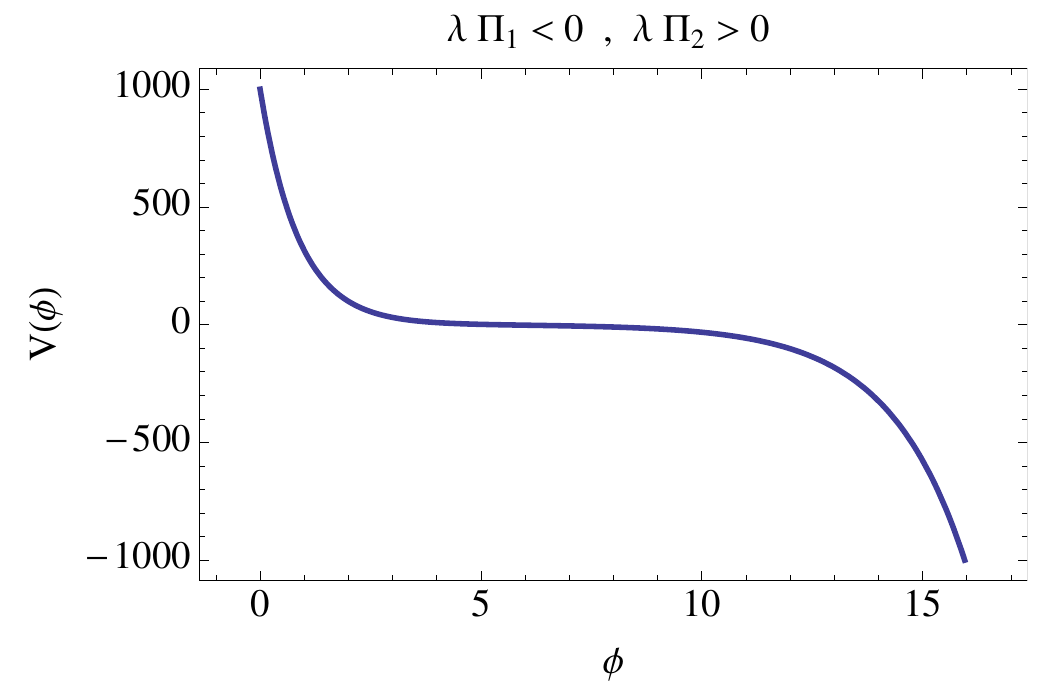}
	\end{minipage}
	\hspace{0.05cm}
\caption{\emph{Scalar potential of Eq. \eqref{firstAttempt} (in units of $g_s^2 \left(\frac{W_0}{\mathcal{V}}\right)^4$) for different choices of the topological integers $\Pi_1$ and $\Pi_2$.}}
	\label{fig:V}
\end{figure}

One may alternatively consider geometries with $\Pi_1<0 \wedge \Pi_2>0$ or $\Pi_1>0 \wedge \Pi_2<0$. The resulting potential will feature an extended plateau where inflation may occur.
Yet in this regime the potential of eq.~\eqref{eq:Vtot} does not have a local minimum and is furthermore unbounded from below. The various possibilities are illustrated in figure \ref{fig:V}. With both $\Pi_i$ negative, the potential becomes a hilltop where inflation may occur on both sides, yet the problem of 
the potential not being bounded from below remains.


In order to obtain sensible inflationary scenarios with a viable exit we now turn our attention to string-loop corrections. 

\subsection{Minima from String Loops}
%
The string-loop corrections to the scalar potential \eqref{gen_potential} for the compactification geometry \eqref{Vol_fibre}, \eqref{2cycle_4cycle} take the form \cite{Cicoli:2008gp}
\begin{align}\label{string_loops}
\delta V_{(g_s)} 
\simeq \frac{g_s |W_0|^2}{\mathcal V^2}\left(g_s^2\frac{(C_1^{KK})^2}{\tau_1^2}+2g_s^2(\alpha C_2^{KK})^2\frac{\tau_1}{\mathcal V^2} \right) \ .
\end{align}
Here we neglected the KK contribution with respect to $\tau_3$, since it carries no dependence on $\tau_1$ and is, thus, not relevant for our discussion. Moreover, note that the winding-mode contributions to $\delta V_{(g_s)}$ are induced by strings which wind one-cycles on the intersection locus of stacks of $D7$-branes. Such one-cycles can exist in the intersection of four-cycles, which the $D7$-branes wrap. However, since the four-cycle associated with $\tau_3$ only intersects with itself, the only winding contribution to $\delta V_{(g_s)}$ that can be induced is via an intersection of the four-cycles associated with $\tau_1$ and $\tau_2$. In the following for simplicity we consider configurations where the $D7$-branes only wrap those four-cycles which are associated with $\tau_2$ or $\tau_3$ and hence winding-mode contributions are completely absent. Note that, in principle we need not assume any special brane-configuration as long as the coefficient of the winding-mode string-loop correction is tuned small enough. 

The string-loop corrections in \eqref{string_loops} are suppressed with additional powers of $g_s^{5/2}$ with respect to $V_{(1)}$ in \eqref{V1correction}. Moreover, noting that the typical size of the topological numbers was inferred in \cite{Ciupke:2015msa} to be $\Pi_i \sim \mathcal{O}(10 \dots 100)$ and using the estimates \eqref{C_gs_estimate} and \eqref{estimate_lambda} we find that 
\begin{equation}\label{estimate_ckk}
 \lvert C_1^{KK} \lvert^2 \sim \lvert C_2^{KK} \lvert^2 \, \ll \, \lvert\lambda\lvert \lvert \Pi_i \lvert \ .
\end{equation}
Thus, for moderately small $g_s \lesssim 10^{-1}$ and $W_0 \gtrsim 1$ the string-loop corrections are suppressed compared to $V_{(1)}$ in some regime of $\tau_1$. However, for small enough and large enough $\tau_1$ the contributions $\delta V_{(g_s),1}^{KK}$ and $\delta V_{(g_s),2}^{KK}$ will become important and eventually dominate over the terms in $V_{(1)}$. Since $\delta V_{(g_s),1}^{KK}$ and $\delta V_{(g_s),2}^{KK}$ are strictly positive, one may use them to stabilise the fibre modulus. In the intermediate $\tau_1$ regime they will remain subleading compared to $V_{(1)}$. In this way one has a scenario where inflation originates from the higher derivative corrections to the scalar potential while a graceful exit results from the interplay between these corrections and string loops. 

The full scalar potential, that is taking into account string loop-corrections and $V_{(1)}$, is now bounded from below and takes the form
\begin{align}\notag\label{FullPotential}
V(\tau_1)&=V^{LVS}(\langle\tau_3\rangle,\langle\mathcal V\rangle)+V_{(1)}+\delta V_{(g_s)}\\ \notag
&=V^{LVS}(\langle\tau_3\rangle,\langle\mathcal V\rangle)-g_s^2 \hat\lambda\frac{{|W_0|}^4}{\mathcal V^4}\left(\Pi_1 \frac{\mathcal V}{\tau_1}+\Pi_2 \lambda_1^{-1/2}\sqrt{\tau_1}\right)\\
&+\frac{g_s |W_0|^2}{\mathcal V^2}\left(g_s^2\frac{(C_1^{KK})^2}{\tau_1^2}+2g_s^2(\alpha C_2^{KK})^2\frac{\tau_1}{\mathcal V^2} \right)+\frac{\epsilon}{\mathcal V^2} \ ,
\end{align}
where an uplift term $\epsilon/\mathcal V^2$ has been added.\footnote{Positive vacuum energy contributions of the form $\epsilon/\mathcal V^p$, $p=1\ldots 3$ can arise in type IIB string compactifications from various sources involving moduli and/or matter F- and/or D-terms, as well as certain anti-branes at the tip of warped throats, see e.g. the discussion on pp. 134 in~\cite{Baumann:2014nda} for an overview with literature.} Upon canonical normalisation $V$ can be recast as
\begin{equation}\label{effectivePotential}
V(\varphi)=V^{LVS}+V_0\left(-\thinspace\mathcal C_1 e^{-2/\sqrt{3}\varphi}-\mathcal C_2 e^{\varphi/\sqrt{3}} +\mathcal C_1^{loop}e^{-4/\sqrt{3}\varphi}+\mathcal C_2^{loop}e^{2\sqrt{3}\varphi} \right)+\delta_{up},
\end{equation}
where we have defined
\begin{align}\notag\label{expressions}
V_0=g_s^2 \frac{{|W_0|}^4}{\mathcal V^4} \thinspace , \quad\mathcal C_1=\hat\lambda\thinspace\Pi_1\mathcal V \thinspace,	\quad \mathcal C_2=\hat\lambda\thinspace\Pi_2\lambda_1^{-1/2} \thinspace,\\ 
\mathcal C_1^{loop}= \frac{\mathcal V^2}{|W_0|^2}g_s(C_1^{KK})^2>0\thinspace,	\quad \mathcal C_2^{loop}= \frac{2g_s}{|W_0|^2}(\alpha C_2^{KK})^2>0
\end{align}
and $\delta_{up}=\epsilon/\mathcal V^2$. We will now tend to the inflationary dynamics in more detail.
\subsection{Inflationary Dynamics}
\begin{figure}[tp]
\centering
\includegraphics[scale=0.6]{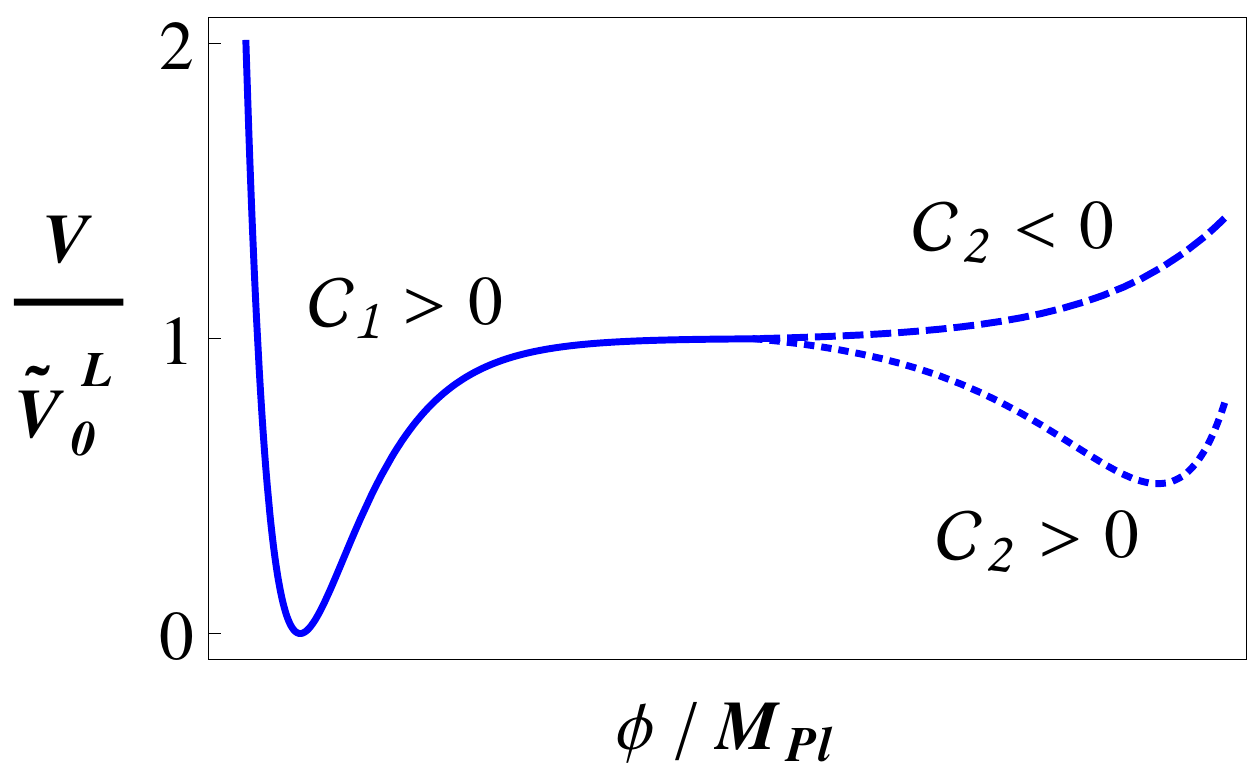}
\includegraphics[scale=0.6]{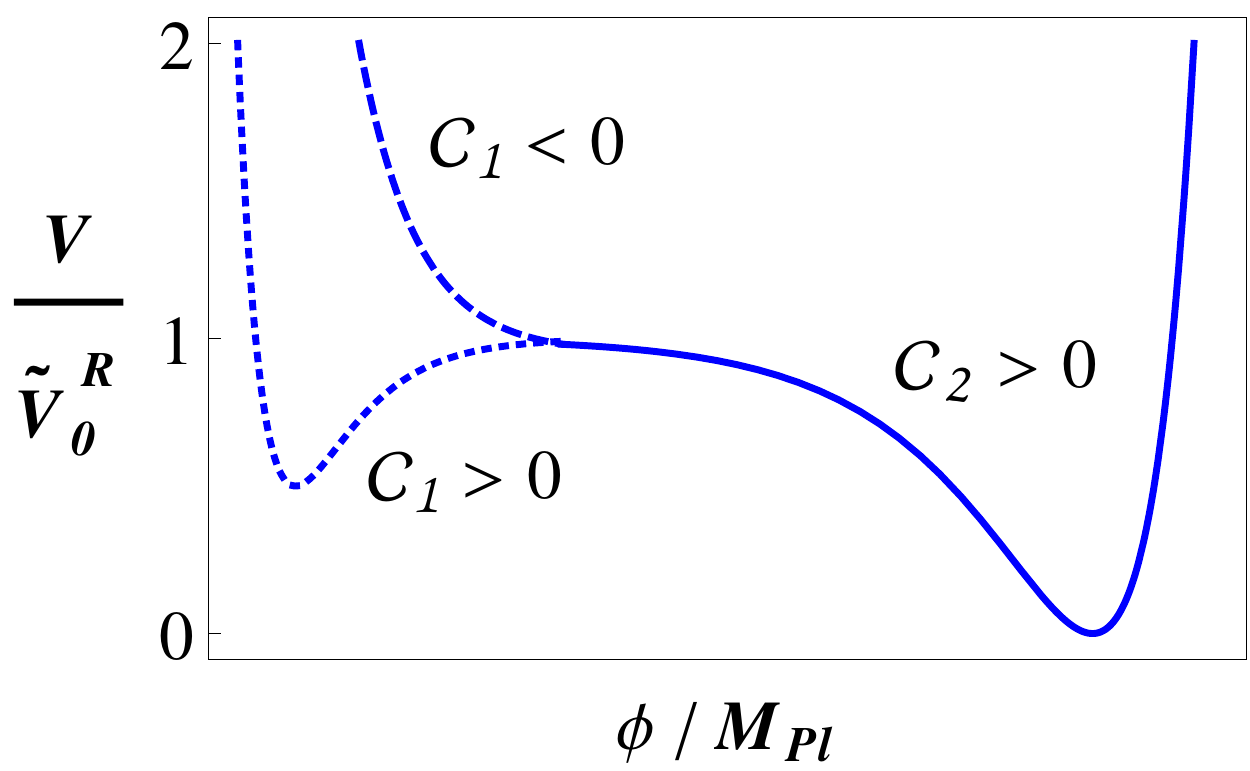}
\caption{\textbf{Left:} \emph{The canonical inflaton potential for $\mathcal C_1>0$ and arbitrary $\mathcal C_2$. String loops ensure that the potential is bounded from below while inflation is driven by the $\mathcal C_1$-term of $V_{(1)}$. If $\mathcal C_2>0$ there can be observed running of the spectral index $n_s$.} \textbf{Right:} \emph{The inflaton rolls to the right. Inflation is driven by the $\mathcal C_2$-term of $V_{(1)}$.}}
\label{potential}
\end{figure}

Since $\mathcal C_1^{loop},\thinspace \mathcal C_2^{loop} > 0$, i.e.\ the string loop corrections will ensure that the potential is bounded from below, the resulting inflationary dynamics depend on the signs of the $\mathcal C_i$. Being initially placed on a suitable side of the minimum, the inflaton can either inflate to the left, the right, or to both sides (hill-top) given appropriate signs of the exponentials coefficients. For both $\mathcal C_i$ negative, the potential will always be dominated by rising exponentials, hence no viable observationally signature appears;
\begin{center}
\begin{tabular}{c|c c c c}
& \emph{left} & \emph{right} & \emph{hill-top} & \emph{exp}\\ 
\hline 
$\mathcal C_1$ & $>0$ & $<0$ & $>0$ & $<0$\\ 
$\mathcal C_2$ & $<0$ & $>0$ & $>0$ & $<0$\\ 
\end{tabular} 
\end{center}
Recalling the relevant terms of the potential during inflation
\begin{equation}\label{PotentialDuringInflation}
V_{inf}\sim -\frac{\mathcal C_1}{\tau_1}-\mathcal C_2\sqrt{\tau_1}\thinspace .
\end{equation}
To ensure a phenomenologically viable inflationary regime, slow roll should be effectively driven by only one of these terms. Thus, we want the observable $\sim 60$ e-folds of inflation to occur before both terms in $V_{inf}$ become of the same order. This happens for values 
\begin{equation}\label{StationaryPoint}
 \tau_1^c \sim \left(\frac{\lvert \mathcal C_1 \lvert}{ \lvert \mathcal C_2  \lvert} \right)^{2/3}\thinspace .
\end{equation}
Now, depending on whether inflation is occurring by the inflaton rolling to the left or right as depicted in figure \ref{potential}, we have to ensure that the plateau appears around values $\tau_1^c > \tau_1$ or $\tau_1^c < \tau_1$.
In order to have the potential sufficiently flat, 
there are bounds on the $\mathcal C_i$ and $\mathcal C_i^{loop}$ coming from the running of the spectral index, which have to be obeyed when considering viable models of inflating to the right or the left. These are subject of subsection \ref{higherorder} and will be found in expression \eqref{runningbound1} and \eqref{runningbound2} respectively.
\subsubsection*{Inflation to the Right}
Let us first consider the case of $\mathcal C_2>0$ and let us take the $\mathcal C_2^{loop}$-term to ensure a minimum; inflation occurs when the field is initially placed on the left side of the minimum and then rolls to the right. This corresponds to an increase of the fibre modulus at constant compactification volume. The inflationary potential is hence given by
\begin{equation}
V_{inf}^R\sim V_0\left(-\mathcal C_2 \sqrt{\tau_1}+\mathcal C_2^{loop}\tau_1 \right)\thinspace,\quad \tau_1>\tau_1^c.
\end{equation}
We find a minimum 
\begin{equation}
\left.\frac{\partial V_{inf}^R}{\partial \tau_1}\right|_{\tau_1>\tau_1^c}\overset{!}{=}0\quad\Rightarrow\quad\tau_1^{min}=\left(\frac{\mathcal C_2}{2\thinspace\mathcal C_2^{loop}} \right)^2,
\end{equation}
and hence
\begin{equation}
V_{inf}^R(\tau_1^{min})=-V_0\frac{\mathcal C_2^2}{4\thinspace \mathcal C_2^{loop}}\equiv -\tilde V_0^R.
\end{equation}
The Minkowski uplifted potential then becomes
\begin{align}\notag\label{VRight}
V_{inf}^R&=\tilde V_0^R\left(1-\beta_R\sqrt{\tau_1}+\frac{\beta_R^2}{4}\tau_1 \right)\\
&=\tilde V_0^R\left(1-\frac{\beta_R}{2}\sqrt{\tau_1} \right)^2\thinspace, \quad \beta_R\equiv \frac{4 \thinspace \mathcal C_2^{loop}}{\mathcal C_2}\thinspace.
\end{align}
Note how the above closes to a square relation. In order to have a sufficient plateau, we require $\beta_R\sqrt{\tau_1}\ll 1\leftrightarrow \tau_1\ll\beta_R^{-2}$, i.e.\ we require $\beta_R\ll1$ to have a plateau at $\tau_1\gtrsim 1$. Recalling expressions \eqref{expressions} for $\mathcal C_2$ and $\mathcal C_2^{loop}$, we see that
\begin{equation}
\beta_R=2\thinspace\frac{g_s^{5/2} ( C_2^{KK})^2}{\lambda\thinspace |W_0|^{2}\Pi_2},
\end{equation}
which may satisfy $\beta_R\ll1$ and hence ensures an inflationary plateau. Further, as we want the minimum to be on the right side of $\tau_1^c$ in \eqref{StationaryPoint}, we require
\begin{equation}
\tau_1^{min}>\tau_1^c\quad\Rightarrow \quad\mathcal C_2^{loop}<\left(\frac{\mathcal C_2^4}{\thinspace |\mathcal C_1|}\right)^{1/3}
\end{equation}
which can easily be satisfied using \eqref{estimate_ckk}. 

Turning to canonical variables via \eqref{canonical_normalization}, we write potential \eqref{VRight} as
\begin{equation}
V^R_{inf}(\varphi)=\tilde V_0^R\left(1-\frac{\beta_R}{2} e^{\frac{\kappa}{2}\varphi}\right)^2
\end{equation}
with $\kappa=2/\sqrt{3}$. Shifting this by the vacuum expectation value $\phi=\varphi+2\thinspace\kappa^{-1}\log(\beta_R/2)$, we arrive at the effective inflaton potential
\begin{equation}\label{VRightCan}
V_{inf}^R(\phi)=\tilde V_0^R\left(1-e^{\frac{\kappa}{2}\phi} \right)^2.
\end{equation}
Finally let us consider the bounds on our parameters arising from the normalisation of the scalar density perturbations \cite{Ade:2015lrj} 
\begin{equation}
\frac{1}{8\pi^2} \frac{H^2}{\epsilon_V}\Big{|}_{*}\sim 2.2\times 10^{-9},
\end{equation}
where the star denotes evaluation at horizon exit which we assume to occur 55 e-folds before the end of inflation. Taking into account the correct normalisation of the potential in the 4D Einstein frame, which brings about an extra $1/8\pi$ factor into $V$ \cite{Conlon:2005ki}, this translates into
\begin{equation}\label{COBE}
\left.\left(\frac{1}{8\pi}\frac{V_{inf}}{\epsilon_V}\right)^{1/4}\right|_{N_e=55}=0.027 \thinspace M_{Pl}
\end{equation}
where $\epsilon_V$ is the potential slow-roll parameter which is independent of the normalisation of the potential. Thus, when inflating to the right, we have
\begin{equation}
\tilde V_0^R=\left. 8\pi \:0.027^4 \thinspace M_P^4 \thinspace\epsilon_V\right|_{N_e=55} =5.7\times 10^{-9},
\end{equation}
where we have set $M_P=1$ in the last equality. This hence sets
\begin{equation}
\tilde V_0^R\equiv  \frac{g_s^2 |W_0|^4}{\mathcal V^4}\frac{\mathcal C_2^2}{4\thinspace \mathcal C_2^{loop}}\sim \lambda^2 \thinspace\frac{|W_0|^6}{\mathcal V^4} g_s^{-2} (C_2^{KK})^{-2}\overset{!}{=}5.7\times 10^{-9}
\end{equation}
which is in accordance with natural choices for $g_s$, $\mathcal V$ and $W_0$.

We may hence summarise the demands and resulting bounds for a viable inflationary regime as follows
\begin{center}
\begin{tabular}{c|c}
\emph{demand} & \emph{resulting bound} \\ 
\hline
plateau at $\tau_1\gtrsim 1$ & $\beta_R\ll 1$ \\ 
$\tau_1^{min}>\tau_1^c$ & $\mathcal C_2^{loop}<\left(\frac{\mathcal C_2^4}{\thinspace |\mathcal C_1|}\right)^{1/3}$ \\ 
COBE & $\lambda^2\thinspace |W_0|^6\thinspace\mathcal V^{-4} g_s^{-2} (C_2^{KK})^{-2}\sim 5\times 10^{-9}$ \\  
\end{tabular} 
\end{center}
If these requirements are met, inflation is fully captured and described by potential \eqref{VRightCan}.

\subsubsection*{Inflation to the Left}

Let us now turn to the case $\mathcal C_1>0$. We ensure the existence of a minimum by considering the $\mathcal C_1^{loop}$-term and hence write the inflationary potential as
\begin{equation}
V_{inf}^L=V_0\left(-\frac{\mathcal C_1}{\tau_1}+\frac{\mathcal C_1^{loop}}{\tau_1^2} \right).
\end{equation}
Being initially placed on the right side of the minimum, inflation now occurs when the inflaton rolls to the left. Geometrically this corresponds to a shrinking fibre and a constant overall volume. The minimum of the potential is now found to lie at
\begin{equation}
\tau_1^{min}=\frac{2\thinspace \mathcal C_1^{loop}}{\mathcal C_1},
\end{equation}
and the value of the potential at the minimum becomes
\begin{equation}
V_{inf}^L(\tau_1^{min})=-V_0\frac{\mathcal C_1^2}{4\thinspace\mathcal C_1^{loop}}\equiv -\tilde V_0^L.
\end{equation}
To be able to safely neglect non-perturbative correction for the Fibre modulus we require that
\begin{equation}
 \tau_1^{min} \gtrsim 1 \ .
\end{equation}
The Minkowski uplifted potential reads
\begin{align}\notag\label{VLeft}
V_{inf}^L&=\tilde V_0^L\left(1-\frac{\beta_L}{\tau_1}+\frac{\beta_L^2}{4\tau_1} \right)\\
&=\tilde V_0^L\left(1-\frac{\beta_L}{2 \tau_1} \right)^2\thinspace,\quad \beta_L=\frac{4\thinspace\mathcal C_1^{loop}}{\mathcal C_1}\thinspace.
\end{align}
Again, the expression closes to a square. We now have $\tau_1^{min}=\beta_L/2$. hence we require $\beta_L\gtrsim 1$ to keep control over the theory. Recall \eqref{expressions} and observe that
\begin{equation}
\beta_L=4 \frac{\mathcal V |W_0|^{-2}g_s^{5/2}(C_1^{KK})^2}{\lambda\thinspace \Pi_1\thinspace }
\end{equation}
which using \eqref{estimate_ckk} and given reasonable choices for $\mathcal{V},g_s$ and $W_0$ guarantees $\beta_L\gtrsim 1$. Also, as inflation now occurs for the field rolling to the left, we seek
\begin{equation}
\tau_1^{min}<\tau_1^c\quad\Rightarrow\quad \mathcal C_1^{loop}<\frac{\mathcal C_1^{5/3}}{2 \, \mathcal C_2^{2/3}} \thinspace, 
\end{equation}
which is easily fulfilled using \eqref{estimate_ckk} and $\mathcal{V} \gg 1$.

Using the canonical normalisation, we rewrite \eqref{VLeft} as
\begin{align}\notag\label{VLeftCan}
V^L_{inf}&=\tilde V_0^L\left(1-\frac{\beta_L}{2}e^{-\kappa\varphi} \right)^2\\
&=\tilde V_0^L\left(1-e^{-\kappa\phi} \right)^2,
\end{align}
where we have shifted the field $\varphi$ to $\phi=\varphi-\kappa^{-1}\log(\beta_L/2)$ with $\kappa=2/\sqrt{3}$.

Again turning to the normalisation of the curvature perturbations \eqref{COBE}, we find the condition
\begin{equation}
\tilde V_0^L=\left.8\pi\: 0.027^4 \thinspace M_P^4 \thinspace \epsilon_V\right|_{N_e=55}=1.5\cdot 10^{-9}.
\end{equation}
or equivalently
\begin{equation}
\tilde V_0^L\equiv  \frac{g_s^2 |W_0|^4}{\mathcal V^4}\frac{\mathcal C_1^2}{4\thinspace\mathcal C_1^{loop}}\sim \lambda^2\thinspace\frac{|W_0|^4}{\mathcal V^4} |W_0|^{2} g_s^{-2}(C_1^{KK})^{-2}\overset{!}{=}1.5\cdot 10^{-9}
\end{equation}
which may be satisfied given reasonable choices of the involved parameters.

We thus summarise the demands and resulting bounds for a viable inflationary regime as follows
\begin{center}
\begin{tabular}{c|c}
\emph{demand} & \emph{resulting bound} \\ 
\hline
minimum at $\tau_1\gtrsim 1$ & $\beta_L\sim g_s^{5/2}\thinspace\mathcal V(C_1^{KK})^2\gtrsim 1$ \\ 
$\tau_1^{min}<\tau_1^c$ & $\mathcal C_1^{loop}<\frac{1}{2}\left(\frac{2}{\mathcal C_2}\right)^{2/3}\mathcal C_1^{5/3}$ \\ 
COBE & $\lambda^2\thinspace|W_0|^6\thinspace\mathcal V^{-4} g_s^{-2} (C_1^{KK})^{-2}\sim 10^{-9}$ \\  
\end{tabular} 
\end{center}

\subsection{Mass Hierarchy}

Up to this point our analysis has focused mostly on the fibre modulus and on the constraints that must be met in order to have a simple single field model with the inflaton rolling either to the left or to the right, towards its minimum. Being a string inspired model, there are however plenty more scalars around, describing various aspects of the geometry of the compactification and the position of branes in the extra dimensions.

In order for the above analysis to hold and for this model to be a genuinely single field inflation model, all the remaining scalars must play no role in the inflationary dynamics. The easiest way to achieve this is for them to be parametrically heavier than the Hubble scale during inflation. The hierarchy in the potential between the large volume part and the inflationary part is crucial for this to happen. More concretely, in this setup we have a total of three complex K\"ahler moduli, various complex structure moduli and the axio-dilaton. These last two get their masses through fluxes which implies
\begin{equation}
m^2_{cs}\sim m^2_S\sim g_s \frac{|W_0|^2}{\mathcal{V}^2}.
\end{equation}
In the K\"ahler moduli sector a mass gap develops as a consequence of the assumed large volume stabilisation, with the blow-up mode $\tau_3$ becoming heavier than the overall volume $\mathcal{V}$:
\begin{equation}
m^2_{\tau_3}\sim g_s\frac{|W_0|^2}{\mathcal{V}^2}\qquad\gg\qquad m^2_{\mathcal{V}}\sim g_s\frac{|W_0|^2}{\mathcal{V}^3}.
\end{equation}
We therefore see that the volume modulus is, apart from the inflaton itself, the lightest field in the spectrum and so the one that would more easily be dynamical during the inflationary epoch. One must therefore make sure that
\begin{equation}
m^2_{\mathcal{V}}\gg H^2 \sim V
\end{equation}
such that no other field besides the fibre modulus $\tau_1$ plays a role in inflation.

For the case when inflation takes place as the fibre modulus rolls to the left, the Hubble scale during the observable inflation is approximately 
\begin{equation}
H^2\sim \tilde{V}_0^L=V_0 \frac{C_1}{\beta_L}\sim g_s^{-2} \frac{W_0^6}{\mathcal{V}^4}\frac{\lambda^2 \Pi_1^2}{(C_1^{KK})^2}
\end{equation}
and so pure single field dynamics requires
\begin{equation}\label{mass_hierarchy_L}
W_0^2\frac{\lambda \Pi_1}{\beta_L}\sim W_0^2 \lambda \Pi_1\ll\sqrt{g_s},
\end{equation}
where we have made use of the fact that in this regime $\beta_L=\mathcal{O}(1)$.

Noting that for inflation towards the right, the Hubble scale is 
\begin{equation}
H^2\sim \tilde{V}_0^R=V_0 \frac{C_2}{\beta_R}\sim g_s^{-2} \frac{W_0^6}{\mathcal{V}^4}\frac{\lambda^2 \Pi_2^2}{(C_2^{KK})^2},
\end{equation}
the requirement of single field dynamics leads to the following bound on the compactification parameters
\begin{equation}\label{mass_hierarchy_R}
\frac{W_0^2}{\mathcal{V}}\lambda \Pi_2 \ll \sqrt{g_s} \beta_R.
\end{equation}
Recall that having all of the inflationary region in the geometric regime ($\tau_1>1$) implies that $\beta_R$ is a small parameter: $\beta_R<1/20$.
\section{Inflationary Observables}\label{observables}

Recall the effective inflationary potentials \eqref{VRightCan} and \eqref{VLeftCan}
\begin{equation}
V_{inf}^R(\phi)=\tilde V_0^R\left(1-e^{\frac{\kappa}{2}\phi} \right)^2\thinspace,\quad V^L_{inf}=\tilde V_0^L\left(1-e^{-\kappa\phi} \right)^2.
\end{equation}
The predictions for a potential of type
\begin{equation}\label{UniversalPotential}
V_{inf}= V_0\left(1-e^{\pm\nu\phi}\right)^2
\end{equation}
for the spectral index $n_s$ and the tensor to scalar ratio $r$ are well approximated by
\begin{align}\notag
n_s&=1-\frac{2}{N}\\
r&=\frac{1}{\nu^2}\frac{8}{N^2}+\mathcal O\left(\frac{1}{N^3} \right),
\end{align}
respectively, where $N$ denotes the number of e-folds before the end of inflation and we have omitted sub-leading terms in $1/N$ for conciseness. The coefficient of the ${\cal O}(1/N^3)$ term can be obtained analytically but we omit its tedious and lengthy form here. Its numerical value differs between our two model classes of 'inflation to the left' and 'inflation to the right', which explains the split in the values for $r$ in Table 1, where the first order phenomenological fingerprint is shown.
\begin{table}[h]
\begin{center}
\begin{tabular}{c|c c c c}
& $n_s(50)$ & $n_s(60)$ & $r(50)$ & $r(60)$\\ 
\hline 
\emph{right} & $0.960$ & $0.967$ & $0.0077$ & $0.0055$\\ 
\emph{left} & $0.960$ & $0.967$ & $0.0024$ & $0.0016$\\ 
\end{tabular} 
\end{center}
\caption{Numerical values for the spectral index $n_s$ and the tensor-to-scalar ratio $r$ calculated at 50 and 60 e-folds before the end of inflation.}
\end{table}

Note how the predictions of inflating to the right are in line with \cite{Cicoli:2008gp}. When inflating to the left however, the tensor to scalar ratio decreases whereas the value of the spectral index remains universal. This is because for 'inflation to the right' the exponential term forming the inflationary plateau is the square-root of the analogous term when inflating to the left.

\subsection{Higher Order Analysis}\label{higherorder}

Recalling the scenarios $\mathcal C_2>0\thinspace,\thinspace \mathcal C_1<0$ or $\mathcal C_1>0\thinspace,\thinspace \mathcal C_2<0$, we see that the resulting inflationary potentials feature an inflection point along the inflationary trajectory. We now turn to study possible higher order corrections to the inflationary observables due to the presence of this inflection point. 

Consider scenario $\mathcal C_2>0\thinspace,\thinspace \mathcal C_1<0$ and the inflationary potential with the $\mathcal C_1$-term included
\begin{equation}
V_{inf}^R\sim V_0\left(-\frac{\mathcal C_1}{\tau_1}-\mathcal C_2\sqrt{\tau_1}+\mathcal C_2^{loop}\tau_1\right).
\end{equation}
The inflaton inflates to the right while the $\mathcal C_1$-term induces the inflection point. Note that we consider the $\mathcal C_1$-term arising from $V_{(1)}$ to be of importance while we have still omitted the string loop induced $\mathcal C_1^{loop}$-term. The reasoning is that first the string loop term is $\tau_1^{-1}$ suppressed with regard to the higher derivative term. Second, for typical values of $C_1^{KK}$ as given in \eqref{C_gs_estimate}, the $C_1^{loop}$-term will be additionally suppressed. At last, while the string loop term scales with $g_s$, the higher derivative term is $g_s^{-3/2}$ enhanced. 

The canonically normalised and uplifted potential hence receives a rising correction at large e-foldings, i.e.\
\begin{equation}
V_{inf}^R(\phi)\sim \tilde V_0^R\left(1-2 e^{\frac{\kappa}{2}\phi}+\varepsilon^2 e^{-\kappa\phi} \right),\quad \varepsilon^2=\frac{|\mathcal C_1|}{4\thinspace \mathcal C_2}\beta_R^3 \thinspace
\end{equation}
where we have already expanded the inflationary potential and have omitted the string loop induced term for notational ease.\footnote{During inflation, the term ensuring the existence of the minimum is negligible.} From the above we may already infer that $\varepsilon^2\ll 1$ for the inflationary plateau not to be spoiled. The slow-roll parameters then receive $\varepsilon^2$ dependent corrections of the form
\begin{align}
\epsilon_V&=\frac{1}{2}\left(\frac{V'}{V}\right)^2=\frac{1}{2}\left(-\kappa\thinspace e^{\frac{\kappa}{2}\phi}-\kappa\thinspace\varepsilon^2 e^{-\kappa\phi}\right)\\
\eta_V&=\frac{V''}{V}=-\frac{1}{2}\kappa^2 e^{\frac{\kappa}{2}\phi}+\kappa^2\varepsilon^2 e^{-\kappa\phi},
\end{align}
where we have taken the potential to be slowly varying during inflation, i.e.\ $V_{inf}^R\sim const.$ Recalling $dN=(2\epsilon_V)^{-1/2}d\phi $, i.e.\
\begin{equation}
N \sim 2 \kappa^{-2} e^{-\frac{\kappa}{2}\phi}+\mathcal O(\varepsilon^2)
\end{equation} 
we hence arrive at the expression for the spectral index $n_s$ including higher order corrections
\begin{equation}\label{nsToRight}
n_s=1-\frac{2}{N}-3\thinspace \varepsilon^2\kappa^4N+\frac{\varepsilon^2\kappa^6}{2} N^2+\ldots
\end{equation}

The above suggests that there is a further phenomenological fingerprint in the form of running of the spectral index which may manifest itself as a loss of power in the CMB temperature power spectrum at large angular scales corresponding to the onset of observable e-folds. Following the analysis of \cite{Broy:2015qna}, we recall that $n_s=d \ln P/ d\ln k =P^{-1} d P/ d N$ and hence
\begin{equation}\label{suppression}
\left.\frac{\Delta P(\delta n_s)}{P}\right|^{N}_{N+\Delta N}=\int\limits_{N+\Delta N}^{N} \delta n_s\thinspace \sim \thinspace\delta n_s\thinspace \Delta N.
\end{equation}
In the case of the inflaton inflating  to the right, we found above that
\begin{equation}
\delta n_s=-3\thinspace \varepsilon^2\kappa^4N+\frac{\varepsilon^2\kappa^6}{2} N^2.
\end{equation}
Taking $N=55$ and requiring $\delta n_s\lesssim 0.008$, which is the 2-$\sigma$ range for the $n_s$ measurement from Planck, we find an upper bound on $\varepsilon^2$ to be \footnote{Note that for $\mathcal{V} \sim 10^3$ and $\Pi_1 \sim \Pi_2$ the below bound requires $\beta_R \lesssim 10^{-3}$, thus, placing a stronger constraint on $\beta_R$ than the minimal required length of the plateau.}
\begin{equation}
\varepsilon^2=\frac{|\mathcal C_1|}{4\thinspace \mathcal C_2}\beta_R^3\thinspace\lesssim \thinspace 2.4\times 10^{-6}.
\end{equation}
Thus, we may now evaluate \eqref{suppression} with the upper bound on $\varepsilon^2$ and the range of e-folds $\Delta N\simeq 5$ over which power-loss is usually considered to occur. We find the power loss to be
\begin{equation}
\left.\frac{\Delta P(\delta n_s)}{P}\right|^{N}_{N+\Delta N} \simeq - 0.04
\end{equation}
which means a suppression of $4$\%.  This is in agreement with earlier works employing exponentially rising corrections to the inflationary plateau \cite{Contaldi:2003zv, Downes:2012gu, Cicoli:2013oba, Pedro:2013pba, Bousso:2013uia, 1404.2278, Kallosh:2014xwa, Cicoli:2014bja, Broy:2014sia}.\footnote{The comparison of our estimate here with the result in~\cite{Pedro:2013pba} may at first looking conflicting. However,~\cite{Pedro:2013pba} wished to ensure a multipole dependence of $\delta n_s$ such that $\delta n_s > 0$ for a drastic blue shift for $\ell < \mathcal O(20)$ and very suddenly reaching $\delta n_s = 0$ at $\ell > \mathcal O(20)$. For such a suddenly shutting down power suppression a much steeper rising exponential is necessary than we get in our models here, as shown in~\cite{Pedro:2013pba}. However, the fit to the data is already consistently improved with a small power loss of $\lesssim 4\%$ which is very slowly shutting down as a function of increasing $\ell$. Such a slowly varying power-loss can arise from more slowly rising exponentials with $\mathcal O(1)$ coefficients in the exponent, as arising in our models here, and hence explaining the difference to~\cite{Pedro:2013pba}.}

Note how the bound on $\varepsilon^2$ also restricts
\begin{equation}\label{runningbound1}
\frac{|\mathcal C_1|}{4 \thinspace\mathcal C_2}\beta_R^3\thinspace\sim\thinspace \lambda^{-3}\thinspace\mathcal V \thinspace g_s^{15/2} \thinspace\Pi_1\thinspace \Pi_2^{-4} (C_2^{KK})^6 \thinspace\lesssim \thinspace 2.4 \times 10^{-6}
\end{equation}
and hence gives a constraint that has to be fulfilled in the first place when considering observationally viable inflation to the right.

Considering the scenario where inflation occurs for the inflaton rolling to the left with $\mathcal C_1>0\thinspace,\thinspace \mathcal C_2<0$, we now start with the potential
\begin{equation}
V_{inf}^L\sim V_0\left(-\frac{\mathcal C_1}{\tau_1}+\frac{\mathcal C_1^{loop}}{\tau_1^2}+\mathcal C_2\sqrt{\tau_1} \right)\thinspace .
\end{equation}
The $\mathcal C_2$-term induces the inflection point. Again, we have omitted the string loop induced term as it is suppressed with regard to the higher derivative $\mathcal C_2$ term by similar reasoning as was employed when justifying the omission of the $\mathcal C_1^{loop}$-term when studying inflation to the right. The canonically normalised and uplifted potential hence receives a rising correction at large e-foldings, i.e.\
\begin{equation}
V_{inf}^L(\phi)\sim \tilde V_0^L\left(1-2 e^{-\kappa\phi}+\varepsilon^2 e^{\frac{\kappa}{2}\phi} \right),\quad \varepsilon^2=\frac{|\mathcal C_2|}{\sqrt{2}\thinspace \mathcal C_1}\beta_L^{3/2}\thinspace
\end{equation}
where we have again expanded the inflationary potential and have omitted the string loop induced term as it plays no role on the inflationary plateau. 
Similarly to the rolling to the right case above, we can derive $n_s$ and get
\beq\label{nsToLeft}
n_s=1-\frac{2}{N}-\frac{3 \sqrt{2} \delta ^2 \sqrt{\kappa ^2
      N}}{N}+\frac{\delta ^2 \kappa ^2 \sqrt{\kappa ^2
      N}}{\sqrt{2}}-\frac{1}{2} 3 \delta ^4 \kappa ^4 N+\ldots\quad.
\eeq
Considering the 2-$\sigma$ bounds by PLANCK, i.e.\ requiring $\delta n_s\lesssim 0.008$ at $N=55$, we obtain the upper bound
\begin{equation}\label{runningbound2}
\varepsilon^2\sim \lambda^{-3/2}\thinspace \mathcal V^{-1}\left(g_s^{5/2}\mathcal V \right)^{3/2}\thinspace\Pi_2\thinspace \Pi_1^{-5/2} (C_1^{KK})^{3/2}\lesssim 10^{-3}\thinspace.
\end{equation}
Saturation of that bound gives again rise to a suppression of power of about $4\%$. Note that the above analysis may also readily be done for both $\mathcal C_i>0$ translating to $\varepsilon^2<0$. The requirement of a sufficiently flat plateau, i.e.\ a constraint on the running of $n_s$, remains unchanged and hence similar bounds are obtained.

Given the bounds on $\varepsilon$ induced by the data-compatible range of $\delta n_s$, we also find that the contribution of the $\delta n_s$ to the magnitude of running $d n_s/d \ln k=-d n_s / d N$ is typically $\lesssim \mathcal O(10^{-4})$ and hence at least an order of magnitude smaller than the contribution to the running from $n_s^{(0)}-1=-2/N$ which is about $d n_s^{(0)}/d \ln k \sim -10^{-3}$.

\subsection{Examples}
In this section we present explicit examples in order to illustrate the points made above on naturalness and tuning of the models under discussion. In table \ref{tab:Examples} we put forth two sets of parameters for each regime.  We assume $\lvert \lambda\lvert =10^{-3}$ throughout, in accordance with the estimate of  Eq. \eqref{estimate_lambda}.
\begin{table}[h]
\begin{center}
\begin{tabular}{c||c|c|c|c|c|c|c|c||c|c}
& $W_0$ & $g_s$ & $\mathcal{V}$ & $\tau_1^{min}$ & $\Pi_1$ & $\Pi_2$ & $C_1^{KK}$& $C_2^{KK}$ & $n_s$ & $r$\\
\hline
\hline
$\mathcal{R}_1$& $5$ & $0.2$& $625.5$& $3000$& $0$& $100$ &$0.00242$& $0.799$& $0.968$&0.0067\\
\hline
$\mathcal{R}_2$& $25$ & $0.3$& $1886.2$& $3500$& $0$& $10$ &$0.000859$& $0.732$& $0.967$&0.0066\\
\hline
$\mathcal{L}_1$& $2$ & $0.3$& $460$& $3$& $100$& $1$ &$0.163$& $0.0288$& $0.966$&0.0018\\
\hline
$\mathcal{L}_2$& $5$ & $0.4$& $1031.6$& $6$& $50$& $0$ &$0.189$& $0.0266$ & $0.969$&0.0021\
\end{tabular}
\end{center}
\label{tab:Examples}
\caption{Examples of compactifications parameters and inflationary observables for inflation to the left ($\mathcal{L}_1$ and $\mathcal{L}_2$) and to the right ($\mathcal{R}_1$ and $\mathcal{R}_2$). Inflationary observables are computed at $N_e=55$.}
\end{table}%

Let us start by discussing some generic features of the inflation to the right regime, corresponding to examples $\mathcal{R}_1$ and $\mathcal{R}_2$. In this regime, control over the $l_s$ expansion in the whole of the  inflationary range leads us to consider large $\tau_1^{min}$.  More concretely, the field excursion between the minimum and the $55$ efoldings point is $\Delta \phi=6.46$, which taking into account the canonical normalisation of Eq. \eqref{canonical_normalization}, implies $\tau_1^{min}\ge 1700$. The overall volume $\mathcal{V}$ is constrained by the normalisation of the density perturbations to be at most $\mathcal{O}(10^3)$. The combination of these two constraints leads to anisotropic compactification manifolds, with a large fibre and a relatively small base, with $\tau_2^{min}\sim \mathcal{O}(10)$. Another stand out feature of this regime is the sensitivity to the topological integer $\Pi_1$. Due to fact that $\Pi_1$ enters in $V$ multiplied by the overall volume,  if one considers manifolds with $\Pi_1\neq0$ the $\mathcal{C}_1$ generated exponential tends to be relevant in the inflationary regime, leading at the very least to deviation from the first order observables of section~\ref{observables} and in more severe cases to the destruction of the inflationary plateau. This situation can be avoided if one considers non-zero $\Pi_1$ and a very large $\Pi_2\sim\mathcal{O}(10^4)$ or alternatively if one considers a manifold with intersection numbers such that $\lambda_1\ll1$. Note that Calabi-Yau threefolds with either $\Pi_1 = 0$ or $\Pi_2 = 0$ can exist. Since we did not choose K\"ahler cone variables, these statement translate into the fact that a particular linear combination of $\Pi'_i$ vanishes, where $\Pi'_i$ encode the topological information of the second Chern class with respect to a basis of cycles spanning the K\"ahler cone as defined in \eqref{def_pi}.

Turning our attention to the inflation to the left regime illustrated by examples $\mathcal{L}_1$ and $\mathcal{L}_2$, we note that these correspond to manifolds with a large base and a relatively small fibre (though examples with larger fibre can be found). Since the exponential generating the inflationary plateau has a coefficient that is volume enhanced with respect to the one that steepens the plateau at large $\phi$,  one can  find points in parameter space that lead to a viable inflationary regime where neither of the topological integers vanishes, while keeping the topological integers in their natural range $\Pi_i\le\mathcal{O}(10^3)$. Once again, the requirement of a mass hierarchy and of the correct normalisation for the amplitude for the density perturbations constrains the overall volume to be $\mathcal{V}\sim \mathcal{O}(10^2-10^3)$. We also note that in accordance to the estimate of Eq. \eqref{C_gs_estimate}, the complex structure dependent coefficients are suitably small: $C_{1,2}^{KK}<1$ in both right and left inflationary regimes.

All in all, these explicit examples show that the inflationary models built in this work are easily obtained in the perturbative regime of both $\alpha'$ and $g_s$ expansions of type IIB orientifold compactifications, without a need for just moderate tuning of the underlying microscopic parameters.

\section{Discussion}\label{discussion}
Knowledge of the string effective action is ever evolving and to date incomplete. Whenever a new work comes about shedding new light into the form of the action, the phenomenologist must investigate what are the  consequences of the novel structures and terms to the history and evolution of the Universe. In this work we have carried out an analysis in this spirit, investigating the consequences of the higher derivative corrections to the action derived in \cite{Ciupke:2015msa} for cosmic inflation in the context of type IIB string compactifications. 

In general terms, the new higher derivative generated terms in the potential are suppressed since they scale as $\mathcal{V}^{-4}$. This renders them harmless and of limited use whenever other larger terms are present. Such terms will therefore have no impact in the K\"ahler moduli stabilisation in the context of LVS on compactification geometries with Swiss-Cheese structure and in the context of KKLT. However, large volume scenarios with different compactification geometries exist in which, through symmetries and cancellations, such terms are the leading ones and are therefore crucial for phenomenology. 

One such setup is analysed here: the stabilisation of the fibre modulus in K3-fibered CY manifolds in the context of LVS compactifications and its inflationary phenomenology. In the absence of non-perturbative effects supported on the fibre modulus, this K\"ahler modulus is flat at leading order in $g_s$ and to third order in $\alpha'$. Previous works have focused on generating a potential for $\tau_1$ via higher order terms in the $g_s$ expansion \cite{Cicoli:2008gp} or via the introduction of non-perturbative effects like poly-instantons  \cite{Cicoli:2011ct}. In this study the potential is generated by the interplay between the higher derivative terms and the string loop corrections to the action. 

Though the higher derivative corrections can by themselves stabilise the fibre modulus, we have shown that the resulting inflationary phenomenology is in tension with the CMB observations. By combining them with the string loop corrections one can simultaneously generate a  controlled minimum and an inflationary plateau. Depending on the underlying microscopic parameters the current model features two regimes which we dubbed inflating to the right and to the left. Both are of Starobinsky-like form yielding $n_s\sim 0.97$ while differing in the tensor fraction: for inflation to the right (left) one finds $r\simeq 0.007$  ($r\simeq 0.002$). 

In our analysis we have not only included the contributions giving rise to the inflationary plateau but also some potentially dangerous terms, whose effect is to spoil the flatness of the inflationary region. We have shown that though not guaranteed at all points in moduli/parameter space, these terms can be rendered harmless for moderately tuned choices of the underlying microscopic parameters. Effectively, the smallish-$g_s$ based decoupling of the string loops from the inflationary plateau, and the large-volume control of successively higher-order $\alpha'$-corrections constitutes a leftover from the approximate K\"ahler moduli shift symmetry at parametrically large volume. However, the leftover is imperfect leading to the still required moderate level of tuning for the string loop coefficients. We have furthermore shown that in some cases such unwanted terms may give rise to a percent level power loss at the level of the primordial scalar power spectrum for which there are some hints in the data.

As a final observation we mention that exponentially flat plateau potentials might also arise from the 3-form flux induced complex structure moduli potential. There, canonically normalizing the kinetic term can produce exponential field dependence in the potential similar to the case of the fibre K\"ahler modulus here. Since in that case the scalar potential would arise at string tree level \emph{and} ${\cal O}(\alpha'^0)$ this might improve the level of protection from higher-order corrections.

\acknowledgments

BB, DC, and AW are supported by the Impuls und Vernetzungsfond of the Helmholtz Association of German Research Centres under grant HZ-NG-603. This work has been supported by the ERC Advanced Grant SPLE under contract ERC-2012-ADG-20120216-320421 and by the grants FPA 2012-32828, and FPA 2010-20807-C02. FGP would also like to thank the support of the Spanish MINECO {\it Centro de Excelencia Severo Ochoa Program} under grant SEV-2012-0249.


\bibliography{alphaprimeinflation}

\end{document}